\documentclass[twocolumn,aps,prb,groupedaddress]{revtex4-2}

\usepackage{graphicx}
\usepackage{color}
\usepackage{amsmath}
\usepackage{enumitem}
\usepackage{amssymb}
\usepackage{hyperref}
\usepackage{cancel}
\usepackage{ulem}
\usepackage{multirow}

\newcommand{\be}{\begin{equation}}
	\newcommand{\ee}{\end{equation}}

\newcommand{\bea}{\begin{eqnarray}}
	\newcommand{\eea}{\end{eqnarray}}

\renewcommand{\vec}[1]{{\boldsymbol #1}}

\begin{document}

\title{
Wide-range $T^2$ resistivity and umklapp scattering in moir\'e graphene 
}

\author{
Hiroaki Ishizuka$^1$, Leonid Levitov$^2$
}

\affiliation{$^1$Department of Physics, Tokyo Institute of Technology, Meguro, Tokyo, 152-8551, Japan.}
\affiliation{$^2$Department of Physics, Massachusetts Institute of Technology,  Cambridge, Massachusetts 02139, USA}

\begin{abstract}
We argue that the unusually strong electron-electron interactions in the narrow bands in moir\'e superlattices originate from compact Wannier orbitals. 
Enhanced overlaps of electronic wavefunctions, enabled by such orbitals, result in a strong el-el superlattice umklapp scattering. We identify the umklapp scattering processes as a source of the strong  
temperature-dependent resistivity observed in these systems. In a simple model, the umklapp scattering predicts a $T$-dependent resistivity that grows as $T^2$ and is getting bigger as the Wannier orbital radius decreases. We quantify the enhancement in el-el scattering by the Kadowaki-Woods (KW) ratio, 
a quantity that is sensitive to umklapp scattering but, helpfully, insensitive to the 
effects due to the high density of electronic states. 
Our analysis predicts anomalously large KW ratio values that clearly indicate the importance of the umklapp el-el processes and their impact on the $T$-dependent resistivity.
\end{abstract}
\date{\today}

\maketitle
%
\vspace{2pc}

\section{Introduction}

Quasiparticle interactions in metals leave unique marks on the temperature ($T$) dependence of transport coefficients, providing a direct way to assess the strength and character of microscopic interactions.
Recent studies of the narrow bands in magic-angle twisted bilayer graphene (TBG)~\cite{dosSantos2007,Li2010,Bistritzer2011}
have revealed a variety of strongly-correlated electronic phases --- Mott-insulating~\cite{Cao2018a}, superconducting~\cite{Cao2018b,Yankowitz2019}, ferromagnetic~\cite{Lu2019,Sharpe2019,Serlin2020}, and nematic~\cite{Choi2019,Kerelsky2019,Jiang2019,Cao2020b}. Simultaneously, a strong $T$-dependent resistivity is observed above the ordering temperatures, growing monotonically and extending to $T$ as high as the bandwidth~\cite{Chung2018,Polshyn2019,Cao2020a}. This resistivity, which is largely insensitive to $T=0$ ordering types, scales as $T$ or $T^2$ depending on the twist angle and carrier density, and is expected to arise from the same electronic interactions as those driving ordering.
A comprehensive study of transport in this system reports both a $T$-linear and a $T^2$ resistivity observed in different parts of phase diagram~\cite{Jaoui2021}.
The $T$-linear resistivity of a Planckian scale is seen in a wide range of moir\'e band occupancies, whereas a $T^2$ resistivity is seen near the band edges, at the low and near-full band occupancies~\cite{Jaoui2021}. 
The $T$-linear behavior attracts intense interest as a possible signature of a strongly-correlated ``strange-metal'' phase~\cite{Homes2004,Zaanen2004,Hartnoll2021}. 
The $T^2$ scaling is suggestive of a Fermi-liquid (FL) picture, however the origin of the observed $T^2$ dependence remains obscure.
Clarifying it will be the aim of this work. 

\begin{figure}[tb]
  \centering
  \includegraphics[width=0.45\linewidth]{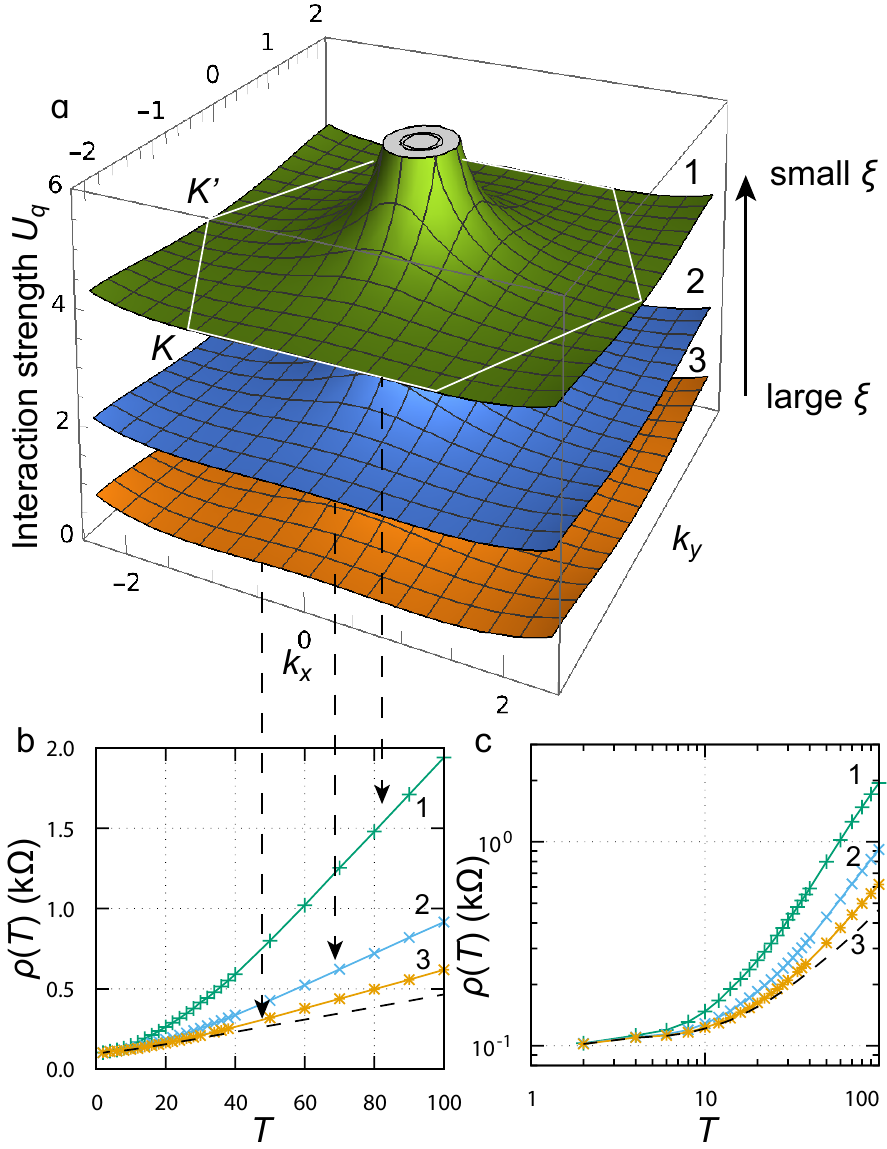}
  \caption{
    (a) Electron interactions in a moir\'e band enhanced by a small Wannier orbital radius.
    Shown is momentum dependence of the el-el interaction, Eq.~\ref{eq:Uq}, for Wannier orbitals of radii $\xi_1=a/6\sqrt3\sim0.10a$ (1; green), $\xi_2=a/3\sqrt3\sim0.19a$ (2; blue), and $\xi_3=a/2=0.5a$ (3; orange), where $a$ is the superlattice parameter. 
    White hexagon marks the superlattice Brillouin zone.
    (b,c) The corresponding $T$-dependent resistivity $\rho(T)=\rho_\text{int}(T)+\rho_\text{imp}(T)$ on a (b) linear and (c) log-log scale.
    Dashed line is the impurity contribution $\rho_\text{imp}(T)$ for free carriers with $v_\text{imp}=0.25$ meV.
  }\label{fig:Uq}
  \vspace{-4mm}
\end{figure}

The two most striking properties of the $T^2$ resistivity reported in Ref.~\cite{Jaoui2021} are its large magnitude and the wide range of temperatures where it occurs, extending from $\sim 20$ K down to deep sub-kelvin temperatures. Namely, it is considerably larger than the resistivity at $T=0$, suggesting a strong microscopic momentum relaxation mechanism that is independent of disorder scattering at $T=0$. As we will see, a strong $T^2$ resistivity that grows without saturation and extends over a wide $T$ range is naturally explained by a mechanism that accounts for the umklapp electron-electron (el-el) scattering by the moir\'e superlattice and an enhancement of these processes due to compact Wannier orbitals. In such processes, the momenta of incoming and outgoing electronic states ($k_{1,3}$ and $k_{2,4}$) satisfy
\begin{eqnarray}
\vec k_1+\vec k_3-\vec k_2-\vec k_4=\vec G\ne0,
\end{eqnarray}
where $\vec G$ is a reciprocal vector of the superlattice.
As is well known, umklapp processes lead to a $T^2$ contribution to the low-temperature resistivity. 
As we argue below, these processes are enhanced by compact Wannier orbitals of small radius $\xi$ that are a staple of TBG bands~\cite{Laissardiere2010,Kang2018,Koshino2018,Carr2019}, in analogy to the recently discussed phonon emission enhancement by electron localization~\cite{Ishizuka2020}. 

The physics of this enhancement is broadly analogous to photon radiation enhancement by mode localization in the Purcell effect in quantum optics.
The enhancement of el-el scattering due to compact Wannier orbitals is a property tunable by the twist angle and other superlattice parameters that control moir\'e bands~\cite{Ishizuka2020}.
As illustrated in Fig.~\ref{fig:Uq}, for a simple model that mimics the key aspects of narrow moir\'e bands, compact Wannier orbitals can produce a considerable enhancement of umklapp-assisted  el-el scattering, leading to resistivity growing in a wide $T$ range.
The resulting $T$ dependence is $T^2$ at the lowest $T$ but, as $T$ grows, it begins to resemble a linear $T$ dependence.

We note that a nonlinear temperature dependence that grows without saturation and extends over a wide range of temperatures is not expected within a FL framework that accounts only for normal (umklapp-free) el-el scattering processes~\cite{Baber1936,Hlubina1995,Rosch1999,Maslov2011,Pal2012}.
Known as the non-Galilean $T^2$ mechanism, it describes el-el scattering in FLs with a non-parabolic band dispersion at $T\ll E_F$.
Because of non-parabolic dispersion, momentum-conserving el-el scattering can lead to current relaxation and ohmic dissipation. However, the resulting $T^2$ dependence quickly saturates due to non-Matthiessen effects arising when the $T^2$ el-el scattering rate exceeds the $T=0$ disorder scattering rate\cite{Maslov2011,Pal2012}. This saturation effect restricts temperature range in which the $T^2$ behavior is predicted by the non-Galilean mechanism. 

It is interesting to mention that many different systems feature resistivity with a large nonlinear $T$ dependence.
These include, in particular, the oxide compounds such as SrTiO$_3$~\cite{Okuda2001,vdMarel2011,Lin2015}, high-$T_c$ cuprates~\cite{Ando2004,Rullier2007}, and Bi$_2$O$_2$Se~\cite{Wang2020} and, notably, a semiconductor with only one Fermi surface~\cite{Lin2015}.
Problems with a common FL explanation has led to considering other possibilities such as the scattering by soft phonons~\cite{Okuda2001,Maslov2017,Zhou2018}.
However, similar to TBG, the origin of the large nonlinear temperature dependence remains poorly understood.

Here, focusing on moir\'e bands, we study how the Wannier functions with small radius $\xi$ affect the large-$q$ part of the el-el interaction, $\tilde U_{\vec q}$, responsible for the el-el umklapp scattering.
As we will see, for $q$ close to the Brillouin zone edge $\tilde U_{\vec q}$ grows rapidly upon reducing $\xi$, giving resistivity that scales as $\rho\propto1/\xi^2$ when $\xi/a\ll1$ (Fig.~\ref{fig:Uq}a).
This yields a large $T^2$ temperature dependence of resistivity reaching 0.5 k$\Omega$ at $T\sim 50$ K for $\theta=1.2^\circ$ TBG.

The FL $T^2$ resistivity mechanism discussed here explains the behavior of resistivity seen near band edges in magic-angle TBG~\cite{Jaoui2021}. In addition, while the ``strange-metal'' linear-$T$ resistivity in TBG conceivably arises from mechanisms that are not part of our FL-based picture, these mechanisms may originate from the strong el-el interactions enabled by compact Wannier orbitals. 
Furthermore, in TBG with non-magic twist angles, the enhanced el-el scattering persists so long as Wannier orbitals remain compact.
In that regard, the key aspect of the spatial structure of Wannier orbitals that boosts umklapp scattering is a compact core, whereas the behavior in the tails -- exponential vs. power-law -- is of lesser importance.
As a result, strong umklapp scattering and $T^2$ resistivity is expected for both the conventional narrow bands and the bands with topological obstructions for truly localized Wannier orbitals such as those discussed in Refs.~\cite{Po2018,Song2021}.
These predictions can be directly tested in slightly non-magic TBG.

The el-el umklapp scattering mechanism explains two key features --- the strong $T^2$ resistivity and its growth up to temperatures comparable to the bandwidth.
To delineate the interaction enhancement due to umklapps from the effects of the high density of states we consider 
the Kadowaki-Woods (KW) ratio~\cite{Rice1968,Kadowaki1986}, a quantity that shows anomalous enhancement reflecting the large umklapp scattering, a convenient way to quantify the el-el scattering strength.

\section{Method}

To analyze the effect of compact Wannier orbitals on the transport properties, we consider a honeycomb model with Wannier function $w({\vec r})\sim e^{-r^2/2\xi^2}$.
This model reproduces the momentum dependence of effect of electron-phonon interaction in TBG~\cite{Ishizuka2020}, and gives a consistent magnitude of resistivity with the experiment~\cite{Polshyn2019}.
The Hamiltonian reads $H=H_0+H_\text{int}+H_\text{imp}$, where 
\begin{eqnarray}\label{eq:hamil}
&& H_{\rm 0}=\sum_{\vec k}
\left(\begin{array}{c}
\psi_{{\vec k},1}^\dagger \\ \psi_{{\vec k},2}^\dagger
\end{array}\right)^{\rm T}
\left(\begin{array}{cc}
0 & h_{\vec k} \\
h_{\vec k}^\ast & 0
\end{array}\right)
\left(\begin{array}{c}
\psi_{{\vec k},1} \\
\psi_{{\vec k},2}
\end{array}\right),\\ 
&& H_\text{imp}=\sum_{\substack{\vec k,\vec k',\alpha,\sigma}} 
\left( \frac{\sqrt3a^2}{2V}\sum_{i}v(\vec r_{i\alpha})e^{-{\rm i}(\vec k-\vec k')\cdot\vec r_{i\alpha}}
\right)
\psi_{\vec k,\alpha}^\dagger\psi_{\vec k',\alpha},
\nonumber\\
&& H_\text{int}=\frac1{2V}\sum_{\substack{\vec k_1,\vec k_2,\vec q,\alpha_{1...4}}}\tilde U_{\vec q}\psi_{\vec k_1,\alpha_1}^\dagger\psi_{\overline{\vec k_1-\vec q},\alpha_2}\psi_{\vec k_2,\alpha_3}^\dagger\psi_{\overline{\vec k_2+\vec q},\alpha_4},\nonumber 
\end{eqnarray} 
where $H_0$ and $H_\text{int}$ are single-particle and interaction terms, respectively. Here, $h_{\vec k}=t\left(1+e^{-i\vec k\cdot\vec a_1}+e^{-i\vec k\cdot\vec a_2}\right)$, $t$ is the nearest-neighbor hopping, ${\vec a}_{1,2}=a\left(\pm\frac12,\frac{\sqrt3}2\right)$, $|\vec a_{1,2}|=a$, are primitive vectors, ${\vec k}=(k_x,k_y)$ is electron momentum, and $V$ is the system volume. 
The single-particle term corresponds to that of the honeycomb lattice model.
The impurity scattering is described by $H_\text{imp}$ where $v(\vec r_{i\alpha})$ is the impurity potential on the site at $\vec r_{ia}$, which is the location of $a$th sublattice in $i$th unit cell.
In the interaction Hamiltonian the matrix element $\tilde U_{\vec q}$ is obtained accounting for the effects of interaction enhancement by the compact Wannier orbitals. Here and below we use a long bar $\overline{\vec k_1-\vec q}$ to denote the vector shifted to the first superlattice Brillouin zone.

\section{Results}

The effective el-el interaction, found from the microscopic interaction averaged over the Wannier-Bloch states, a simple model that mimics the essential ingredients of the TBG problem, takes the form
\begin{eqnarray}
\tilde U_{\vec q}=\frac{e^2}{2\epsilon}\sum_{\vec G}\frac{e^{-\xi^2(\vec q+\vec G)^2/2}}{\sqrt{(\vec q+\vec G)^2+q_{TF}^2}},
\label{eq:Uq}
\end{eqnarray}
where $q_{TF}^{-1}$ is the Thomas-Fermi screening length. 
This interaction matrix is obtained for an extended Hubbard-like model with Coulomb interactions and Gaussian Wannier functions of radius $\xi$ (see \ref{sec:Uq}).
As illustrated in Fig.~\ref{fig:Uq}a, the quantity $\tilde U_{\vec q}$ grows rapidly upon reducing $\xi$.

To study the effect of $\xi$ on the resistivity, we calculate the resistivity using the variational method for Boltzmann theory~\cite{Ziman2001}.
Assuming the conventional form of the field-induced modulation of electron distribution $\delta f_{n\vec k}= f_{n\vec k}-f_{n\vec k}^0=\tau e\vec E\cdot\vec v_{n\vec k}\beta f^0_{n{\vec k}}(1-f^0_{n{\vec k}})$~\cite{Ishizuka2021}, the resistivity follows Mathiessen's rule $\rho(T)=\rho_\text{int}(T)+\rho_\text{imp}(T)$, where
\begin{eqnarray}
\rho_\text{int}(T)=&
\frac{\sum\limits_{k_i,s_i} Q^\text{(int)}_{1234} \left[v^x_{\vec{k}_1s_1}+v^x_{\vec{k}_3s_3}-v^x_{\vec{k}_2s_2}-v^x_{\vec{k}_4s_4}\right]^2}{4k_BTe^2 \left\{\int\frac{dk^2}{(2\pi)^2}(v_{\vec{k}s}^x)^2\frac{\partial f_{\vec{k}s}^0}{\partial \varepsilon_{\vec{k}s}}\right\}^2}\label{eq:ziman}
\end{eqnarray}
is the interaction contribution, with $\sum_{k_i,s_i}$ a shorthand for $\sum_{s_i}\int\prod_i\frac{dk_i^2}{(2\pi)^2}$. Here we introduced notation
\begin{eqnarray}\label{eq:Q1234}
Q^\text{(int)}_{1234}&=&\frac{(2\pi)^3}\hbar \left(\frac{1}{4} \tilde U_{\overline{\vec k_1 -\vec k_2}}\right)^2
M_{1234} f_{\vec{k}_1s_1}^0f_{\vec{k}_3s_3}^0(1-f_{\vec{k}_2s_2}^0)\nonumber\\
&&\times (1-f_{\vec{k}_4s_4}^0) \delta(\overline{\vec{k}_1+\vec{k}_3-\vec{k}_2}-\vec{k}_4)\nonumber\\
&&\times 
\delta(\varepsilon_{\vec{k}_1s_1}+\varepsilon_{\vec{k}_3s_3}-\varepsilon_{\vec{k}_2s_2}-\varepsilon_{\vec{k}_4s_4})
.
\nonumber
\end{eqnarray}
Here, $\varepsilon_{\vec k s}=s|h_{\vec k}|$ and $v^x_{\vec k s}=\partial_{k_x}\varepsilon_{\vec k s}/\hbar$ are respectively the eigenenergy and the group velocity of the electron with momentum $\vec k$ and band index $s=\pm$, $f_{\vec ks}^0=1/(e^{(\varepsilon_{\vec ks}-\mu)}+1)$ is the Fermi distribution function, $\mu$ is the chemical potential, $\hbar$ is the Planck constant, and the sum over $\vec G$ is that for the reciprocal lattice vectors. The quantity $M_{1234}=[1+s_1s_2s_3s_4\cos(s_1\phi_{\vec{k}_1}-s_2\phi_{\vec{k}_2}+s_3\phi_{\vec{k}_3}-s_4\phi_{\vec{k}_4})]/2$ originates from the coherence factors for the two-body scattering matrix elements.

Similarly, the impurity contribution to resistivity is
\begin{eqnarray}
&\rho_\text{imp}(T)= \frac{\sum_{s,s'}\int\frac{dk^2}{(2\pi)^2}\frac{dk'^2}{(2\pi)^2}\,\tilde{Q}^\text{(imp)}_{\vec{k}s;\vec{k}'s'}\left[v_{\vec{k}'s'}^x-v_{\vec{k}s}^x\right]^2}{2k_BTe^2 \left\{\sum_s\int\frac{dk^2}{(2\pi)^2}(v_{\vec{k}s}^x)^2\frac{\partial f_{\vec{k}s}^0}{\partial \varepsilon_{\vec{k}s}}\right\}^2},\label{eq:rhoimp}\\
&\tilde{Q}^\text{(imp)}_{\vec{k}s;\vec{k}'s'}=\frac{\sqrt3}2\frac{\pi v^2a^2}{\hbar}f_{\vec{k}s}^0(1-f_{\vec{k}s}^0)\delta(\varepsilon_{s,\vec{k}}-\varepsilon_{s',\vec{k}'}),
\end{eqnarray}
We assume the random average of the impurity potential to be $\langle v^{(\alpha)}_{\vec q}(v^{(\beta)}_{\vec q'})^\ast\rangle=\frac{v_\text{imp}^2}{N}\delta_{\alpha\beta}\delta_{\vec q\vec q'}$.

Fig.~\ref{fig:Uq}b shows the temperature dependence of $\rho(T)$ for a bandwidth $W$ ($=3t$) corresponding to $\theta=1.2^\circ$ twist angle.
In the calculation, we used dielectric constant $\varepsilon=30\varepsilon_0$ inferred from recent theoretical estimates~\cite{Goodwin2019,Vanhala2020}.
Here, $\varepsilon_0$ is the dielectric constant of the vacuum.
The solid lines are $\rho(T)$ for three different radii of Wannier orbitals  $\xi_1<\xi_2<\xi_3$, the dashed line is $\rho_\text{imp}(T)$.
The weak $T$ dependence of $\rho_\text{imp}(T)$, nonvanishing even at $k_BT\ll W$, originates from a variation of the density of states at the Fermi level.
The interaction resistivity significantly increases by reducing $\xi$ exceeding the impurity resistivity, confirming the enhancement of resistivity by quenched Wannier orbitals.
The $T^2$ behavior survives up to $\sim 50$ K for $\theta=1.2^\circ$, to a temperature comparable to the bandwidth.

The monotonic increase of  $\rho(T)$ without saturation is in stark contrast to the behavior of the $T^2$ resistivity predicted by the non-Galilean mechanism. In this case the $T^2$ behavior occurs at the lowest temperatures and saturates when the el-el scattering rates become comparable to the disorder scattering rate~\cite{Maslov2011,Pal2012}.
In the case of umklapp scattering, the non-Mathiessen correction by the normal scattering merely renormalizes the prefactor of $T^2$ resistivity but does not result in its saturation~\cite{Maebashi1998}.

\begin{figure}[tb]
  \centering
  \includegraphics[width=0.8\linewidth]{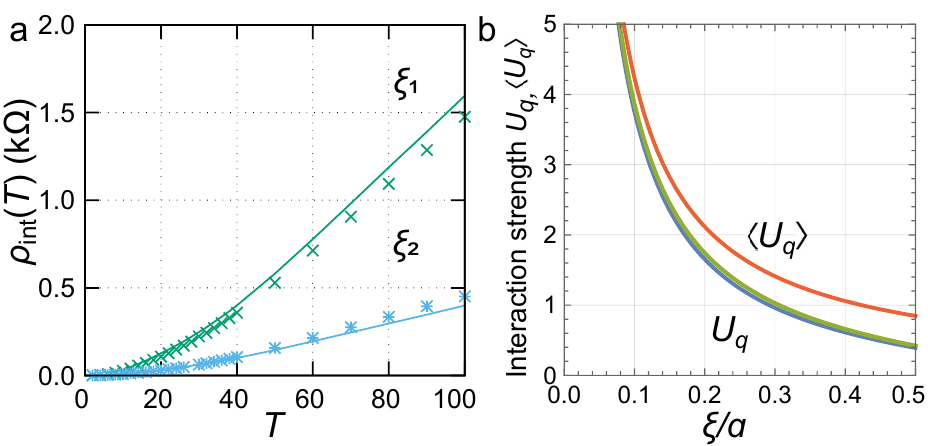}
  \caption{
    Comparison of the interaction $\tilde U_{\bf q}$, Eq.\ref{eq:Uq}, and the $q$-independent model $\langle\tilde U_{\vec q}\rangle$, Eq.\ref{eq:Uq_ave}. 
    (a) Resistivity $\rho_\text{int}(T)$ for $\xi=\xi_1$ and $\xi_2$ at filling $\nu=2/3$ calculated from the full interaction $\tilde U_{\bf q}$ (crosses and stars) and $\langle\tilde U_{\vec q}\rangle$ (solid lines).
    The bandwidth and lattice parameter are $W=13.4$ meV and $a=11.7$ nm, which corresponds to twisted bilayer graphene with twist angle $\theta=1.2^\circ$.
    (b) Comparison of $\langle \tilde U_{\vec q}\rangle$ and $\tilde U_{\vec q}$ at $K$ point. Blue and green lines are obtained for different Thomas-Fermi screening lengths $q_{TF}^{-1}=\sqrt3a$ and $10^2\sqrt3a$, respectively.
  }\label{fig:UqvsV0}
\end{figure}

The enhancement in the el-el interaction due to compact Wannier orbitals and umklapp scattering translates into an anomalously large Kadowaki-Woods (KW) ratio~\cite{Rice1968,Kadowaki1986}.
This quantity is defined as $\alpha=\rho_\text{int}(T)/C_e^2(T)$ where $C_e(T)$ is the electron specific heat.
The KW ratio is $T$-independent at small $T$ and is normalized by the density of states at the Fermi level.
As such, itprovides a convenient metric allowing to delineate the effects of el-el scattering enhancement due to umklapp scattering and due to the large density of states at the Fermi level.

For example, for 2D electrons with quadratic dispersion the resistivity reads $\rho(T)\propto (U/\epsilon_F)^2n^2(T/\epsilon_F)^2$ whereas $C_e(T)\propto nT/\epsilon_F$ at the low temperatures; here, $\varepsilon_F$ is the Fermi energy and $n$ is the electron density.
Hence, $\alpha\sim (U/\varepsilon_F)^2$ is a temperature-independent quantity that characterizes the strength of the el-el scattering in ``natural units''~\cite{Coleman2016}.
Empirically, this ratio is $\alpha\sim10^{-5}\;\mu\Omega$ cm (mJ mol$^{-1}$ K$^{-1}$)$^{-2}$ in transition-metal compounds~\cite{Rice1968}, and $\alpha\sim10^{-6}\;\mu\Omega$ cm (mJ mol$^{-1}$ K$^{-1}$)$^{-2}$ in heavy-fermions, showing similar values for materials in the same group.
An exception is high-$T_c$ cuperates, where it is $\alpha\sim10^{-2}\;\mu\Omega$ cm (mJ mol$^{-1}$ K$^{-1}$)$^{-2}$ in experiment~\cite{Proust2016}.

In our 
problem, at $T=5$ K the 
heat capacity for the model in Eq.~\ref{eq:hamil} is $C_e(T)=0.56$ mJ mol$^{-1}$ K$^{-1}$ for 1 mol carbon atoms, where we multiplied $\rho_\text{int}(T)$ and $C_e(T)$ by 1/4 and 4, respectively, 
to account for the four-fold spin/valley degeneracy. 
This gives the KW ratio $\alpha=1.3\;\mu\Omega$ cm (mJ mol$^{-1}$ K$^{-1}$)$^{-2}$ for $\xi=\xi_1=a/6\sqrt3$ and $\alpha=3.4\times10^{-1}\;\mu\Omega$ cm (mJ mol$^{-1}$ K$^{-1}$)$^{-2}$ for $\xi=\xi_2=a/3\sqrt3$, where the results were normalized to the thickness of two graphene monolayers, $c=6.7$ \AA, to allow for a direct comparison to 
3D materials.
The resulting values, anomalous for 
typical materials, are as high as those for high-$T_c$ cuprates~\cite{Proust2016} and quasi-2D oxides~\cite{Hussey2005}.

To further examine the effect of the large-$q$ scattering, we compare the resistivity with that calculated using a $q$-independent scattering rate model.
We focus on the $q_{TF}\to0$ limit and approximate $(\tilde U_{\vec q}^2/4)^2$ by a constant $\langle \tilde U_{\vec q}\rangle^2/4$, where
\begin{eqnarray}\label{eq:Uq_ave}
\langle \tilde U_{\vec q}\rangle
\sim&\frac{\sqrt3e^2a^2}{16\pi^2\epsilon}\int dq^2 \frac{e^{-\xi^2|\vec q|^2/4}}{|\vec q|}=\frac{\sqrt3}{8\sqrt\pi}\frac{e^2a^2}{\epsilon\xi}.
\end{eqnarray}
is the average over the first Brillouin zone, which should be acceptable when $\xi/a$ is small because $U_{\vec q}$ is almost constant except for the peak at the $\Gamma$ point, as in Fig.~\ref{fig:UqvsV0}b.
In Fig.~\ref{fig:UqvsV0}a, we show $\rho_\text{int}(T)$ calculated using $\tilde U_{\vec q}$ and that using $\langle \tilde U_{\vec q}\rangle^2$.
The results show good agreement implying that the large-$q$ part plays the key role in the resistivity.
In the constant scattering-rate approximation, the resistivity scales $\langle \tilde U_{\vec q}\rangle^2\propto \xi^{-2}$, which explains the significant increase of $\rho_\text{int}$ with decreasing $\xi$ (Fig.~\ref{fig:Uq}b).

\begin{figure}[tb]
  \centering
  \includegraphics[width=0.8\linewidth]{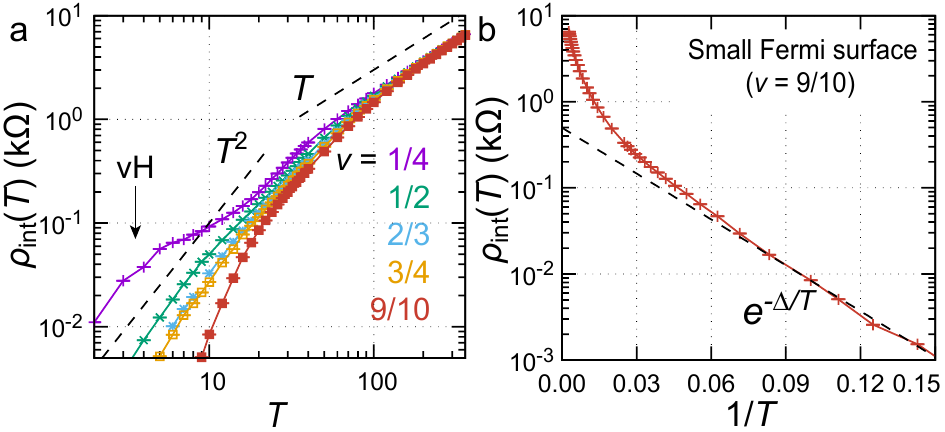}
  \caption{
    The contribution to resistivity due to  el-el scattering, $\rho_\text{int}(T)$.
    (a) Log-log plot of the resistivity for several fillings $\nu$ derived from $\langle\tilde U_{\vec q}\rangle^2$. Dashed lines are $T^2$ and $T$.
    Enhanced low-$T$ resistivity at $\nu=1/4$ originates from van Hove singularity.
    (b) Resistivity for a small Fermi surface $k_F<G/4$ ($\nu=0.9$). The umklapp scattering is exponentially suppressed at low temperatures, giving $\rho_\text{int}(T)$ that displays an activation-type temperature dependence. Dashed line: $\exp(-2|\mu-\mu_0|/k_BT)$  with $2|\mu-\mu_0|/k_B=40.9$ K.
    The bandwidth and lattice parameter are respectively $W=13.4$ meV and $a=11.7$ nm, which corresponds to twisted bilayer graphene with twist angle $\theta=1.2^\circ$.
  }\label{fig:ndep}
\end{figure}

The temperature dependence of $\rho_\text{int}$ for different fillings $-1<\nu<1$ is shown in Fig.~\ref{fig:ndep}a.
Here, we define $\nu=0$ as a filling at which the Fermi level is at the Dirac point. Accordingly, for $\nu=+1$ ($\nu=-1$) the band is fully filled (empty).
As discussed above, at $k_BT\ll W$ the resistivity scales as $\rho_\text{int}(T)\propto T^2$, whereas at $k_BT\gtrsim W$ linear scaling is found, $\rho_\text{int}(T)\propto T$.
To understand the origin of this behavior, we focus on temperatures $k_BT\gg W$. In this case, we can approximate Fermi distributions and their derivatives appearing in Eqs.~\ref{eq:ziman} and \ref{eq:Q1234} as 
\begin{eqnarray} \nonumber
&&f_{\vec ks}\sim(\nu+1)/2, \quad 1-f_{\vec ks}\sim(1-\nu)/2, 
\\ \nonumber
&&\frac{\partial f_{\vec{k}s}^0}{\partial \varepsilon_{\vec{k}s}}=-f_{\vec{k}s}^0(1-f_{\vec{k}s}^0)/k_BT\sim(1-\nu^2)/(4k_BT).
\end{eqnarray}
In this limit,  Eq.~\ref{eq:ziman} predicts linear scaling 
\begin{eqnarray}
\rho_\text{int}(T)\propto T
\end{eqnarray}
with a prefactor that does not depend on $\nu$.
This $\nu$ and $T$ dependences differ from those for $\rho_\text{imp}(T)$ and the resistivity due to electron-phonon scattering $\rho_\text{ph}(T)$:
a similar calculation for Eq.~\ref{eq:rhoimp} gives $\rho_\text{imp}\propto T/(1-\nu^2)$, and that for $\rho_\text{ph}(T)$ is $\rho_\text{ph}\propto T^2/(1-\nu^2)$~\cite{Ishizuka2021}.
The $T$ and $\nu$ dependences 
can be directly compared to those found in TBG, where the filling $\nu$ can be tuned arbitrarily throughout the narrow bands.

It is interesting to mention that the $T^2$ contribution to $\rho_\text{int}(T)$ due to umklapp scattering is quenched at small $T$ when the Fermi surface becomes small enough. Here we illustrate this behavior for $\nu=9/10$.
In general, the umklapp scattering does not contribute to the resistivity when Fermi wavenumber is as small as $k_F<G/4$.
Under these conditions, the resistivity, rather than vanishing, is exponentially suppressed, 
\begin{eqnarray}
\rho_\text{int}(T)\propto e^{-2(\mu_{T=0}-\mu^\ast)/k_BT}
,
\end{eqnarray}
 where $\mu_{T=0}$ is the chemical potential at $T=0$ and $\mu^\ast$ is the chemical potential for the Fermi wavenumber $k_F=G/4$. This is so because the only electrons contributing to the resistivity are those thermally excited to a large momentum state that feature umklapp scattering.
The exponential suppression makes the umklapp resistivity negligible in conventional semiconductors.
For TBG, in contrast, the narrow bandwidth yields $(\mu_{T=0}-\mu^\ast)/k_BT$ values that are relatively small, allowing the umklapp scattering to contribute to resistivity.
In particular, in Fig.~\ref{fig:ndep}, $(\mu_{T=0}-\mu^\ast)/k_B\sim20$ K for $\nu=9/10$.
Accordingly, the umklapp resistivity increases as $\rho_\text{int}(T)\propto e^{-2(\mu_{T=0}-\mu^\ast)/k_BT}$ as the temperature grows, where $\mu_{T=0}$ is the chemical potential at $T=0$ [Fig.~\ref{fig:ndep}b].
As shown in Fig.~\ref{fig:ndep}a, however, the temperature dependence at moderate $T$ does not look much different from that for fillings away from band edges, i.e. for a large Fermi surface.
We therefore expect a substantial superlinear temperature dependence due to umklapp scattering to be observable close to the band edges and charge neutrality point.

\section{Conclusions}

In summary, compact Wannier orbitals in moir\'e bands can lead to strong superlattice el-el umklapp scattering.
As argued above, the scattering rate grows as $\propto1/\xi^2$ upon the Wannier orbital radius $\xi$ decreasing.
The enhanced umklapp scattering manifests itself through a large $T^2$ resistivity and an anomalously large KW ratio. 
The essential requirement for this mechanism is a compact spatial structure of Wannier orbitals, however, spatial localization is not required.
Therefore the predicted enhancement of el-el umklapp scattering, the large $T^2$ resistivity and the anomalous KW ratios are expected to occur for both non-topological bands and the bands that host topologically nontrivial electronic states such as those described in Ref.~\cite{Po2018,Song2021}.
Narrow bands in TBG is a system where recent electronic state calculations predict compact Wannier orbitals~\cite{Laissardiere2010,Kang2018,Koshino2018,Carr2019}.
At the same time, transport measurements in TBG report strong temperature-dependent resistivity, $T$-linear in the strange metal phase and $T^2$ in the FL phase~\cite{Jaoui2021}, pointing to a key role of the el-el interactions enhanced by compact Wannier orbitals. 


We thank D. Bandurin and D. Efetov for inspiring discussions.
This work was supported by JSPS KAKENHI (Grant Numbers JP18H03676, JP19K14649), by the Science and Technology Center for
Integrated Quantum Materials, NSF Grant No. DMR1231319 and Army Research Office Grant W911NF-18-1-0116.

\appendix
\begin{widetext}

\section{Electron-electron interaction $\tilde U_{\vec q}$}\label{sec:el-el}
\label{sec:Uq}

\subsection{The el-el interaction Hamiltonian in the Wannier orbital representation}

We consider an interaction term,
\begin{eqnarray*}
H_{ee}=\frac12\sum_{\sigma,\sigma'}\int d^2r'd^2r\,\psi_\sigma^\dagger(\vec{r})\psi_\sigma(\vec{r})U(\vec{r}-\vec{r}')\psi_{\sigma'}^\dagger(\vec{r}')\psi_{\sigma'}(\vec{r}').
\end{eqnarray*}
Here, $U(\vec{r})$ is the electron-electron interaction and $\psi_\sigma(\vec{r})$ [$\psi_\sigma^\dagger(\vec{r})$] are the annihilation (creation) operators of an electron with spin $\sigma$ at position $\vec{r}$. To project the interaction to the low-energy states, we approximate the electron operators by
\begin{eqnarray*}
\psi_\sigma(\vec{r})\sim\sum_{i,a} w(\vec{r}-\vec{R}_i^a)c_{a\sigma,i},
\end{eqnarray*}
where $w(\vec{r})$ is a Wannier function (orbital) centered at $\vec{r}=\vec0$ and $c_{a\sigma,i}$ is the annihilation operator for the state $w(\vec{r})$ centered at $\vec{R}_i^a\equiv \vec{R}_i+\vec{r}^a$. Here, $i$ and $a$ are respectively the indices for the unit cell and the orbital; $\vec{R}_i$ is the position of $i$th unit cell and $\vec{r}^a$ is the relative position of $a$th orbital within a unit cell.
After a Fourier tranform, $c_{a\sigma,i}=\frac1{\sqrt{N}}\sum_{\vec k}e^{{\rm i}\vec k\cdot\vec R_i}c_{a\sigma,\vec k}$, where $c_{a\sigma,\vec k}$ is the annihilation operator for electron with momentum $\vec k$, orbital $a$, and spin $\sigma$, the interaction Hamiltonian reads
\begin{eqnarray*}
&& H_{ee}=\frac12\sum_{\substack{\sigma,\sigma',a_j,\vec{k}_j}}\int\frac{dq^2}{(2\pi)^2}g^{a_1\sigma,\vec{k}_1,a_2\sigma,\vec{k}_2}_{a_3\sigma',\vec{k}_3,a_4\sigma',\vec{k}_4}(\vec{q})\,c_{a_1\sigma,\vec{k}_1}^\dagger c_{a_2\sigma,\vec{k}_2} c_{a_3\sigma,\vec{k}_3}^\dagger c_{a_4\sigma,\vec{k}_4},\\
&&g^{a_1\sigma,\vec{k}_1,a_2\sigma,\vec{k}_2}_{a_3\sigma',\vec{k}_3,a_4\sigma',\vec{k}_4}(\vec{q})=\sum_{i_j}\frac{U_{\vec{q}}}{N^2}\int d^2r'd^2r\,
e^{-{\rm i}\vec{q}\cdot(\vec{r}-\vec{r}')+{\rm i}\vec{k}_1\cdot\vec{R}_{i_1}-{\rm i}\vec{k}_2\cdot\vec{R}_{i_2}+{\rm i}\vec{k}_3\cdot\vec{R}_{i_3}-{\rm i}\vec{k}_4\cdot\vec{R}_{i_4}}\\
&&\times w^\ast(\vec{r}-\vec{R}_{i_1}^{a_1})w(\vec{r}-\vec{R}_{i_2}^{a_2})w^\ast(\vec{r}'-\vec{R}_{i_3}^{a_3})w(\vec{r}'-\vec{R}_{i_4}^{a_4}).
\end{eqnarray*}
The interaction matrix $g^{a_1\sigma,\vec{k}_1,a_2\sigma,\vec{k}_2}_{a_3\sigma',\vec{k}_3,a_4\sigma',\vec{k}_4}(\vec{q})$ consists of two parts: diagonal terms $\vec R_{i_1}^{a_1}=\vec R_{i_2}^{a_2}$ and $\vec R_{i_3}^{a_3}=\vec R_{i_4}^{a_4}$, and off-diagonal terms $\vec R_{i_1}^{a_1}\ne\vec R_{i_2}^{a_2}$ or $\vec R_{i_3}^{a_3}\ne\vec R_{i_4}^{a_4}$.
The diagonal terms corresponds to extended Hubbard interactions $c_{a\sigma i}^\dagger c_{a\sigma i}c_{a'\sigma' j}^\dagger c_{a'\sigma' j}$, namely a density-density coupling.
On the other hand, the off-diagonal terms are small in the case of a small Wannier orbital radius $\xi\ll|\vec R_{i_1}^{a_1}-\vec R_{i_2}^{a_2}|$ or $\xi\ll|\vec R_{i_3}^{a_3}-\vec R_{i_4}^{a_4}|$ because of small overlaps of Wannier orbitals with different centers.

\subsection{Hubbard-like terms}

As we are interested in the small Wannier orbital radius limit, we only consider the Hubbard-like interactions, $i_1=i_2$, $a_1=a_2$ and $i_3=i_4$, $a_3=a_4$.
The terms omitted in this approximation involve overlaps of different Wannier orbitals.
Hence, in the limit of compact Wannier orbitals $\xi\ll a$, these terms are expected to be much smaller than the Hubbard-like terms.
Our numerical estimates show that they are indeed small and therefore safe to ignore.

Within the Hubbard-like approximation, $g_{\vec{q}}(a_1\sigma,\vec{k}_1,a_2\sigma,\vec{k}_2;a_3\sigma',\vec{k}_3,a_4\sigma',\vec{k}_4)$ reads
\begin{eqnarray*}
g^{a_1\sigma,\vec{k}_1,a_2\sigma,\vec{k}_2}_{a_3\sigma',\vec{k}_3,a_4\sigma',\vec{k}_4}(\vec{q})=U_{\vec{q}}|\eta_{\vec{q}}|^2\delta_{\overline{\vec{k}_1-\vec{k}_2-\vec{q}},\vec{0}}\delta_{\overline{\vec{k}_3-\vec{k}_4+\vec{q}},\vec0},
\end{eqnarray*}
where the form factor $\eta_{\vec{q}}=\int d^2r\,|w(\vec{r})|^2e^{{\rm i}\vec{q}\cdot\vec{r}}$, contains the information of Wannier function.
As in the main text, the overline $\overline{\vec k+\vec q}$ denotes wavevectors shifted by a reciprocal lattice vector into the first Brillouin zone.
This notation is used to relate the extended zone scheme and the folded zone scheme used in our analysis.
Using this interaction term, the interaction part $H_{ee}$ reads
\begin{eqnarray*}
H_{ee}=\frac12\sum_{\substack{\sigma,\sigma',a_j,\vec{k}_j\in1BZ}}\int\frac{dq^2}{(2\pi)^2}U_{\vec{q}}|\eta_{\vec{q}}|^2\,c_{a_1\sigma,\vec{k}_1}^\dagger c_{a_1\sigma,\overline{\vec{k}_1-\vec{q}}} c_{a_3\sigma',\vec{k}_3}^\dagger c_{a_4\sigma',\overline{\vec{k}_3+\vec{q}}},\\
\qquad=\frac12\sum_{\substack{\sigma,\sigma',a_j,\vec{k}_j\in1BZ}}\int_{1BZ}\frac{dq^2}{(2\pi)^2}\tilde{U}_{\vec{q}}\,c_{a_1\sigma,\vec{k}_1}^\dagger c_{a_1\sigma,\overline{\vec{k}_1-\vec{q}}} c_{a_3\sigma',\vec{k}_3}^\dagger c_{a_3\sigma',\overline{\vec{k}_3+\vec{q}}},\\
\tilde{U}_{\vec{q}}=\sum_{\vec{G}}U_{\vec{q}}|\eta_{\vec{q}+\vec{G}}|^2.
\end{eqnarray*}
Here, the sum over $k_j$ is over momenta in the first Brillouin zone and the sum over $\vec G$ is over all reciprocal vectors.
The result shows that the Wannier function form factor renormalizes the Hubbard interaction as $U_{\vec{q}}\to\tilde{U}_{\vec{q}}$. 

The discrete version of $H_{ee}$ reads
\begin{eqnarray*}
H_{ee}=&\frac1{2V}\sum_{\substack{\sigma,\sigma',a_j,\vec{k}_j,\vec{q}\in1BZ}}\tilde{U}_{\vec{q}}\,c_{a_1\sigma,\vec{k}_1}^\dagger c_{a_1\sigma,\overline{\vec{k}_1-\vec{q}}} c_{a_3\sigma',\vec{k}_3}^\dagger c_{a_3\sigma',\overline{\vec{k}_3+\vec{q}}}.
\end{eqnarray*}

\subsection{Gaussian Wannier functions and Yukawa interaction}

As an illustration, we consider Gaussian Wannier orbitals and a Yukawa interaction model.
The Gaussian model is a simple example of compact Wannier orbital and Yukawa potential is the effective interaction between electrons after taking account of Thomas-Fermi screening.
The corresponding direct-space and momentum-space expressions read
\begin{eqnarray*}
w(\vec{r})=\frac1{\sqrt\pi\xi}e^{-\frac{r^2}{2\xi^2}},\qquad \eta_{\vec{q}}=e^{-(\xi^2q^2/4)},
\end{eqnarray*}
and
\begin{eqnarray*}
U(\vec{r})=\frac{e^2}{4\pi\varepsilon r}e^{-q_{TF}r},\qquad U_{\vec{q}}=\frac{e^2}{2\varepsilon}\frac{1}{\sqrt{q_{TF}^2+q^2}}.
\end{eqnarray*}
Using these quantities, the interaction matrix $\tilde{U}_{\vec{q}}$ reads
\begin{eqnarray*}
\tilde{U}_{\vec{q}}=&\frac{e^2}{2\varepsilon}\sum_{\vec{G}}\frac1{\sqrt{q_{TF}^2+|\vec{q}+\vec{G}|^2}}e^{-\xi^2|\vec{q}+\vec{G}|^2/2}e^{{\rm i}(\vec{q}+\vec{G})\cdot(\vec{r}_{a_3}-\vec{r}_{a_1})}.
\end{eqnarray*}
This is the interaction matrix considered in the main text, Eq.~\ref{eq:Uq}.

For a small $q$ near the $\Gamma$ point, $\tilde{U}_{\vec{q}}$ is dominated by $G=0$ term, $\tilde{U}_{\vec{q}}\propto1/\sqrt{q_{TF}^2+q^2}$, and depends strongly on $q$.
On the other hand, $\tilde{U}_{\vec{q}}$ near the Brillouin zone edge is almost independent of $\vec q$, as shown in Fig.~\ref{fig:Uq}. This justifies the constant interaction-matrix approximation $\tilde{U}_{\vec{q}}\sim\langle\tilde{U}_{\vec{q}}\rangle$ used in the analysis summarized in Figs. 2 and 3 of the main text.
\end{widetext}

%

\vspace{1cm}

\end{document}